# Three-dimensional energy gap and origin of charge-density wave in kagome superconductor KV$_3$Sb$_5$

Takemi Kato,[1,#] Yongkai Li,[2,3,#] Tappei Kawakami,[1,#] Min Liu,[2,3,#] Kosuke Nakayama,[1,4,*] Zhiwei Wang,[2,3,*] Ayumi Moriya,[1] Kiyohisa Tanaka,[5,6] Takashi Takahashi,[1,7,8] Yugui Yao,[2,3] and Takafumi Sato[1,7,8,9,*]

[1]*Department of Physics, Graduate School of Science, Tohoku University, Sendai 980-8578, Japan*

[2]*Centre for Quantum Physics, Key Laboratory of Advanced Optoelectronic Quantum Architecture and Measurement (MOE), School of Physics, Beijing Institute of Technology, Beijing 100081, China*

[3]*Beijing Key Lab of Nanophotonics and Ultrafine Optoelectronic Systems, Beijing Institute of Technology, Beijing 100081, China*

[4]*Precursory Research for Embryonic Science and Technology (PRESTO), Japan Science and Technology Agency (JST), Tokyo, 102-0076, Japan*

[5]*UVSOR Synchrotron Facility, Institute for Molecular Science, Okazaki 444-8585, Japan*

[6]*School of Physical Sciences, The Graduate University for Advanced Studies (SOKENDAI), Okazaki 444-8585, Japan*

[7]*Center for Science and Innovation in Spintronics, Tohoku University, Sendai 980-8577, Japan*

[8]*Advanced Institute for Materials Research (WPI-AIMR), Tohoku University, Sendai 980-8577, Japan*

[9]*International Center for Synchrotron Radiation Innovation Smart (SRIS), Tohoku University, Sendai 980-8577, Japan*

#These authors equally contributed to this work.

*e-mail: k.nakayama@arpes.phys.tohoku.ac.jp; zhiweiwang@bit.edu.cn; t-sato@arpes.phys.tohoku.ac.jp

**Abstract**

Kagome lattices offer a fertile ground to explore exotic quantum phenomena associated with electron correlation and band topology. The recent discovery of superconductivity coexisting with charge-density wave (CDW) in the kagome metals KV$_3$Sb$_5$, RbV$_3$Sb$_5$, and CsV$_3$Sb$_5$ suggests an intriguing entanglement of electronic order and superconductivity. However, the microscopic origin of CDW, a key to understanding the superconducting mechanism and its possible topological nature, remains elusive. Here, we report angle-resolved photoemission spectroscopy of $KV_3Sb_5$ and demonstrate a substantial reconstruction of Fermi surface in the CDW state that accompanies the formation of small three-dimensional pockets. The CDW gap exhibits a periodicity of undistorted Brillouin zone along the out-of-plane wave vector, signifying a dominant role of the in-plane inter-saddle-point scattering to the mechanism of CDW. The characteristics of experimental band dispersion can be captured by first-principles calculations with the inverse star-of-David structural distortion. The present result indicates a direct link between the low-energy excitations and CDW, and puts constraints on the microscopic theory of superconductivity in alkali-metal kagome lattices.

**Introduction**

Kagome lattice is at the forefront of exploring exotic quantum states owing to its peculiar geometry characterized by a two-dimensional (2D) network of corner-sharing triangles. The representative geometrical effect appears as quantum magnetism in insulating kagome lattice, where the strong magnetic frustrations inherent in the triangular coordination lead to a quantum spin liquid [1, 2]. In the metallic counterparts, the electronic states originating from the kagome-lattice symmetry are of particular interest, as they consist of a nearly flat band, Dirac-cone band, and saddle-point van Hove singularity which often dominate the physical properties of strongly correlated electron systems and topological materials. When the electron filling is tuned for the flat band, ferromagnetism [3] or charge fractionalization [4, 5] may appear, whereas the tuning of Dirac-cone band is expected to create strongly correlated Dirac semimetal, topological insulator, and Weyl semimetal phases [6-9]. Recent experiments reported several model kagome materials with the properly tuned electron-filling and the realization of predicted exotic properties [10-21]. On the other hand, the electron filling at the saddle point in kagome materials has been scarcely realized despite several intriguing theoretical predictions such as unconventional density wave orders [6, 22-24], superconductivity [23-27], and nematic instability [24].

Recently, $AV_3Sb_5$ (A = K, Rb, and Cs; see Fig. 1a for crystal structure) [28-30] has emerged as the first kagome compound suitable for studying the physics associated with the saddle-point band. $AV_3Sb_5$ commonly undergoes a charge-density wave (CDW) transition at $T_{CDW}$ = 78-103 K, accompanied by the in-plane unit-cell doubling with the 2×2 periodicity [31-34] and additional out-of-plane doubling (2×2×2 CDW) [33, 35] or quadrupling (2×2×4) [36]. This CDW shows an intriguing entanglement with the superconductivity (transition temperature $T_c$ of 0.9-2.5 K) [28-30] and non-trivial topological surface states [29]. Also, while a static magnetic order is absent [28-30], a strong anomalous Hall effect [37, 38] and a possible time-reversal-symmetry-breaking CDW state [31, 39-42] have been reported, highlighting the unconventional nature of CDW in $AV_3Sb_5$.

Despite accumulating experimental and theoretical studies [31-36, 39-60], the origin of CDW in $AV_3Sb_5$ is highly controversial. A fundamental issue is the type of structural distortion responsible for the CDW formation. This is essential to unveil the mechanism of superconductivity, because the superconductivity appears in the distorted phase. First-principles calculations proposed two types of distortions sharing the same space group of *P6/mmm* (No. 191) to account for the in-plane 2×2 periodicity (Fig. 1a).

One is the "Star-of-David" (SoD) distortion of V atoms which has a close connection with a well-known motif of a strongly correlated CDW state in transition-metal dichalcogenides [61]. The other is an inverse type of the SoD distortion, where V atoms show an opposite displacement compared with the SoD case, resulting in a periodic arrangement of triangular and hexagonal patterns, called "Tri-Hexagonal" (TrH; i.e., inverse SoD) structure. Although both phases are energetically more stable than the undistorted (1×1) phase in the calculations [54], it is experimentally highly controversial which distortion actually takes place [31-36, 53]. Besides the type of distortion, it is also unclear how the distortion influences the electronic states and how the 3D nature of CDW manifests itself in the electronic states.

In this study, we provide insights into these key questions through the investigation of low-energy excitations in full **k** space by utilizing photon-energy-tunable ARPES. We demonstrate the appearance of a 3D pocket due to CDW-induced electronic reconstruction and an anisotropic CDW gap maximized around the saddle point of kagome V band. These characteristics can be reproduced by first-principles calculations assuming TrH structural distortion, and further suggest the importance of inter-saddle-point scattering for the occurrence of CDW.

**Results and Discussion**

**Fermi surface reconstruction by CDW**

At first, we present the Fermi surface (FS) topology of $KV_3Sb_5$. Figure 1c shows the ARPES-intensity mapping at $E_F$ as a function of $k_x$ and $k_y$ at $T = 120$ K (above $T_{CDW}$) measured with 114-eV photons which probe the $k_z \sim 0$ plane of the bulk Brillouin zone (purple shade in Fig. 1b; see Supplementary Fig. 1 for the relationship between $h\nu$ and $k_z$ in $KV_3Sb_5$). A circular pocket centered at the Γ point and two (small and large) triangular shaped intensity patterns centered at each K point are resolved, as also visualized in Fig. 1d. According to the band structure calculations (Supplementary Fig. 2), they are attributed to the $5p_z$ band of Sb atoms embedded in the kagome-lattice plane (Sb1 in Fig. 1a) and the kagome-lattice band with mainly the $3d$ character of V atoms (V in Fig. 1a) [31, 35], respectively (note that the small and large triangular features have the dominant $d_{x^2-y^2}$ and $d_{xz/yz}$ character, respectively). The large triangular feature with the V-$3d_{xz/yz}$ character connects to each other around the M point and forms a large hexagonal FS centered at the Γ point. One can also identify bright spots around the M point which originate from the large density of states associated with the saddle-point van Hove singularity in the band dispersion. The observed FS topology is consistent with previous ARPES reports of $AV_3Sb_5$ and is well reproduced by the density-functional-theory (DFT) calculations [29, 37, 43-47, 49, 50].

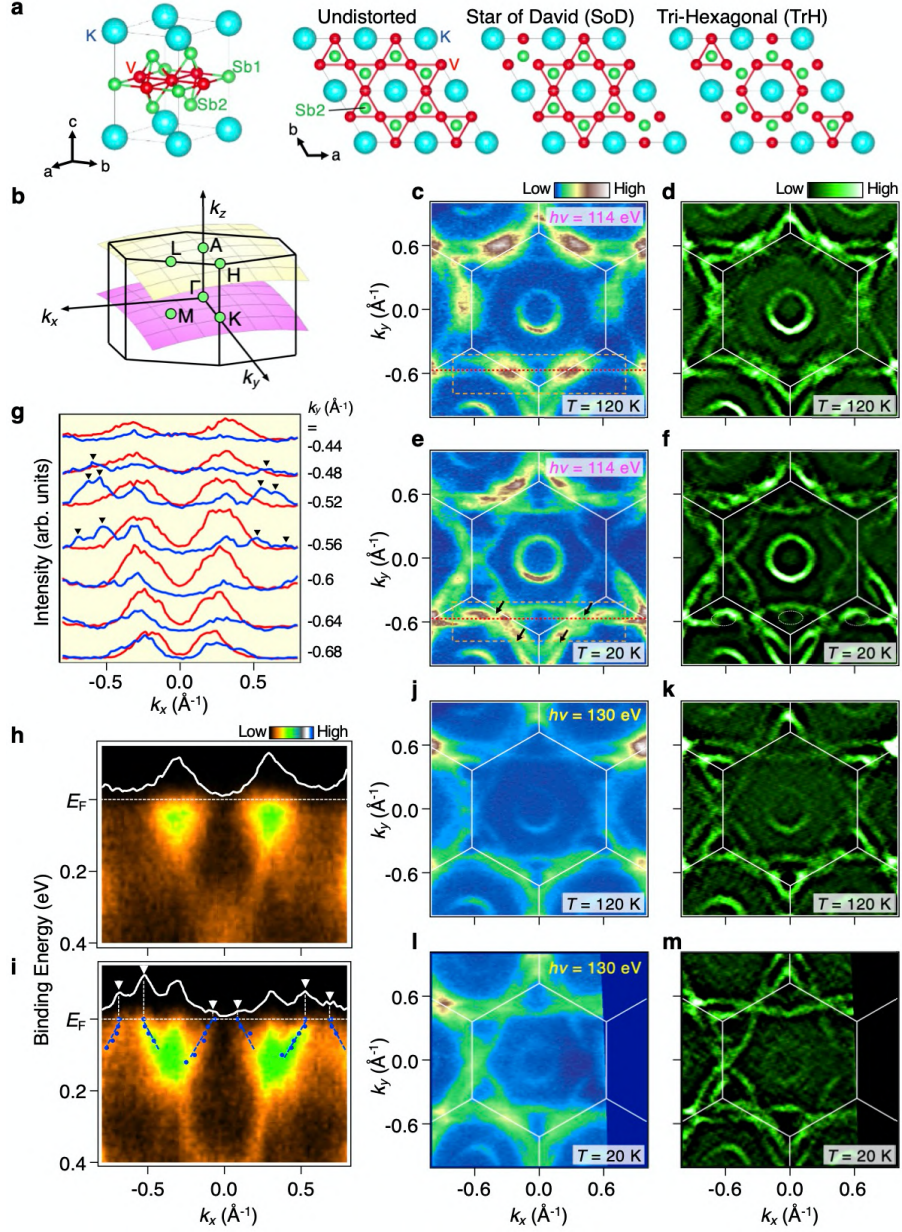

**FIG. 1**. **Reconstruction of Fermi surface in $KV_3Sb_5$. a,** Crystal structure of $KV_3Sb_5$. Kagome-lattice layers without structural distortion, with the Star-of-David (SoD) and Tri-Hexagonal (TrH) structural distortions are also indicated. Shorter V-V bonds are higlighted by red lines. **b,** Bulk Brillouin zone of $KV_3Sb_5$ together with the $k$ planes covered by the measurements at $h\nu$ = 114 and 130 eV. **c, d** ARPES-intensity maps at $E_F$ plotted as a function of $k_x$ and $k_y$, measured at $T$ = 120 K at $h\nu$=114 eV (corresponding to $k_z \sim 0$ plane; purple shade in **b**), and corresponding second-derivative plot of ARPES intensity, respectively. **e, f** Same as **c** and **d** but measured at $T$ = 20 K. Black arrows in **e** highlight the electronic reconstruction. White dashed circles in **f** are a guide for the eyes to trace the reconstructed small pockets. **g,** Comparison of MDCs at $E_B = E_F$ at 120 K (red) and 20 K (blue), measured in the $k$ region enclosed by orange rectangle in **c** and **e**. Black triangles show the MDC peaks originating from reconstructed Fermi surfaces below $T_{CDW}$. **h, i** ARPES intensity measured at $k_y$ = -0.57 Å$^{-1}$ (~M-M cut, shown by a red dotted line in **c** and **e**) at $T$ = 120 K and 20 K, respectively. Blue dots and lines are a guide for the eyes to trace the reconstructed band dispersion (see Supplementary Fig. 3). White curve in the inset shows the MDC at $E_F$. White triagnles show MDC peaks originating from the folded bands. **j-m,** Same as **c-f** but measured at $h\nu$ = 130 eV (corresponding to $k_z \sim \pi$ plane; yellow shade in **b**).

Now we turn our attention to the influence of CDW on the FS. While ARPES intensity mappings at $T$ = 20 K and 120 K (Figs. 1c and 1e) share several common features such as the Γ-centered electron pocket and the triangular pattern around the K point, a closer look reveals some intrinsic differences between them. For example, the intensity around the M point associated with the saddle point is substantially suppressed at $T$ = 20 K due to the CDW-gap opening. Also, the intensity of triangular pockets is strongly distorted at $T$ = 20 K to show a discontinuous behavior at particular **k** points (black arrows), in contrast to that at $T$ = 120 K which shows a smooth intensity distribution. This indicates the reconstruction of FS due to the strong modulation of band dispersions by the periodic lattice distortion associated with the CDW, which has not been resolved in previous ARPES studies of $AV_3Sb_5$ [29, 37, 43-47, 49, 50]. Plot of second-derivative

ARPES intensity at $T = 20$ K in Fig. 1f signifies that the discontinuous intensity distribution (Fig. 1e) is accompanied with the emergence of a small pocket-like feature near the K point (white dotted ellipse). This pocket is associated with the CDW because it is absent at $T = 120$ K (Fig. 1d). The appearance/absence of the pocket-like feature below/above $T_{CDW}$ is more clearly seen from a direct comparison of momentum distribution curves (MDCs) at several $k_y$ slices in Fig. 1g; the peaks in MDCs marked by triangles, which correspond to the pocket-like feature, appear only below $T_{CDW}$. Therefore, our observation supports the Fermi-surface reconstruction due to CDW.

To obtain further insights into the FS reconstruction, we look at the change in the band dispersion across $T_{CDW}$. Figures 1h and 1i show a comparison of the ARPES intensity along a **k** cut indicated by a red dashed line in Figs. 1c and 1d which traverses the reconstructed FSs at $T = 20$ K. One can recognize an obvious difference in the intensity distribution between $T = 120$ K and 20 K in Figs. 1h and 1i. A new holelike band which crosses $E_F$ (indicated by blue circles and lines) appears at $T = 20$ K in the **k** region where FS is absent at $T = 120$ K, as also identified from the MDC at $E_F$ (top panels; see white triangles; also see a comparison of MDCs in Supplementary Fig. 3). This confirms the emergence of a small hole-pocket-like feature around the K point due to the CDW-induced band folding (Supplementary Fig. 3d). It is noted that some reconstructed Fermi surfaces predicted by calculations are not clearly resolved in the present study, possibly because of their weak intensity due to the matrix-element effect in the ARPES measurement.

To clarify the three-dimensional (3D) electronic states above $T_{CDW}$, we have also mapped out the FS at $k_z \sim \pi$ by using $h\nu = 130$ eV photons, and show the ARPES intensity and corresponding second-derivative plots at $T = 120$ K in Figs. 1j-1m. While the overall intensity at $k_z \sim \pi$ (Fig. 1j) is similar to that at $k_z \sim 0$ (Fig. 1c) as seen in the existence of a circular pocket at the $\bar{\Gamma}$ point and triangular FSs at the $\bar{K}$ point, the bright intensity around the M point seen at $k_z = 0$ (Fig. 1c) is absent at $k_z \sim \pi$ (Fig. 1j). This is because the saddle-point band is located well above $E_F$ at $k_z \sim \pi$ [29, 35, 37, 43] due to the $k_z$ dispersion. To see the influence of CDW at $k_z \sim \pi$, we have also mapped out the ARPES intensity at $T = 20$ K (Figs. 1l and 1m), and found no clear FS reconstruction as highlighted by the absence of small pocket-like feature around the $\bar{K}$ point. This behavior is different from that at $k_z \sim 0$ at $T = 20$ K (Figs. 1e and 1f), suggesting that the pocket-like feature associated with the CDW-induced FS reconstruction observed at $k_z \sim 0$ does not form a 2D cylindrical FS in the 3D Brillouin zone but has a 3D ellipsoidal shape.

**Three dimensionality of CDW gap**

Now that the CDW-induced FS reconstruction is established, next we investigate the CDW gap. At $T = 120$ K, the ARPES intensity and corresponding EDCs in Figs. 2c and 2d, respectively, measured along the MK cut (cut 1 in Fig. 2a) signify the $E_F$ crossing of the holelike V-$3d_{x^2-y^2}$ band (black circles; called SP1). This band is connected to the electronlike dispersion along the $\Gamma$M cut [29, 35, 46, 48] to form a saddle point slightly above $E_F$ at the M point. There is another holelike band with the V-$3d_{xz/yz}$ character (called SP2) which forms another saddle point near $E_F$. At $T = 20$ K, the SP1 band does not cross $E_F$, but stays below $E_F$ at $E_B$ of 80-100 meV (Fig. 2e) due to the CDW-gap opening (see Supplementary Fig. 4 for details about the numerical fitting of the band dispersion). This is also visible from the EDCs in Fig. 2f and consistent with the intensity suppression around the M point seen in Fig. 1e. On the other hand, the SP2 band does not show such a large CDW gap and stays near $E_F$. These observations demonstrate that the most prominent change in the energy bands across $T_{CDW}$ occurs in the SP1 band. When the **k** cut is chosen so as to pass slightly away from the M point (cut 2 in Fig. 2a), the SP1 band (black circles) moves closer to $E_F$ and stays at $E_B$ = 50-80 meV (Figs. 2g and 2h), signifying that the magnitude of CDW gap is sensitive to the in-plane wave vector $k_{//}$. Along cut 3 which passes the electron pocket at Γ (Figs. 2i and 2j), the band crosses $E_F$ with no gap opening, suggesting that this FS pocket (Sb 5$p$ orbital) is not a main player of CDW. To elucidate the **k** dependence of CDW gap, we measured ARPES data along various **k** cuts parallel to the cuts 1-3 at $T$ = 20 K and extracted EDCs at the **k** points where the leading edge shows a smallest shift relative to $E_F$ (called minimum gap locus [62]) along each cut. The result plotted in Fig. 2k shows that the gap gradually becomes smaller on approaching the Γ point (cut 3; $|\Delta k_x|$ = 1) from the M point (cut 1; $|\Delta k_x|$ = 0). This is also seen from the Lorentzian fitting of the EDCs in Fig. 2l that displays a clear energy shift of the EDC peak toward higher binding energy around the M point as opposed to the peak at $E_F$ around Γ (compare the peak positions of the black curves). It is noted that, to obtain a reasonable fit to the EDC around the M point, we assumed the presence of in-gap state (the green curve in the bottom panel of Fig. 2l) which may originate from, e.g., CDW-induced folded bands (also see Supplementary Figs. 5a-5c). The magnitude of CDW gap shown in Fig. 2m visualizes its strong **k** dependence with the maximum around the M point. With the maximum gap size $\Delta$ of ~80 meV, the ratio $\Delta/k_B T_{CDW}$ is calculated to be ~12. Although the accurate estimation of the coupling constant $2\Delta/k_B T_{CDW}$ (where $2\Delta$ corresponds to the full gap) is difficult because the energy gap is likely particle-hole asymmetric, the obtained large ratio clearly indicates that the CDW in KV$_3$Sb$_5$ is in the strong coupling regime. This implies the unconventional nature of the CDW.

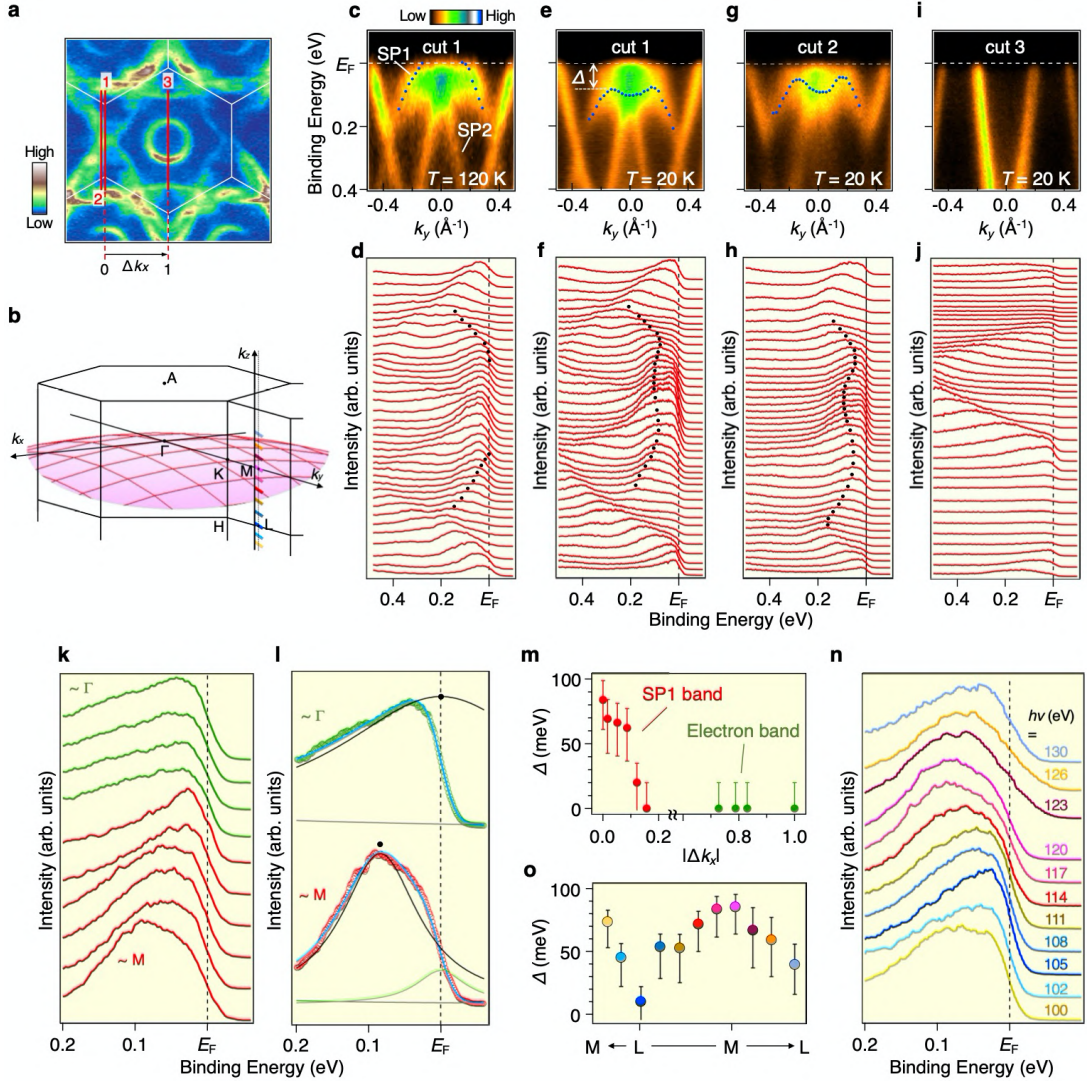

**FIG. 2. Momentum dependence of CDW gap in 3D Brillouin zone. a**, ARPES intensity map at $E_F$ at $h\nu$ = 114 eV ($k_z \sim 0$ plane; same as Fig. 1e), together with **k** cuts where high-resolution ARPES measurements were performed (dashed red curves: cuts 1-3). $|\Delta k_x|$ is defined as the $k_x$ value at the measured **k** point/cut relative to that at the M point. The ΓM length was set to be a unity. **b**, Bulk Brillouin zone of $KV_3Sb_5$. Purple shade and colored horizontal lines indicate the $k_z = 0$ plane and the **k** cuts where the EDCs shown in **n** were obtained, respectively. **c, d** ARPES intensity and corresponding EDCs along cut 1 measured at $T$ = 120 K. **e, f** Same as **c** and **d** but measured at $T$ = 20 K. **g-j** Same as **e** and **f** but measured along cut 2 (**g** and **h**) and cut 3 (**i** and **j**). Blue and black dots in **c-h** show experimental band dispersions obtained by tracing the peak position in EDCs; note that the dots in **c** and **d** were obtained by tracing the peak position in EDCs divided by the Fermi-Dirac distribution function at $T$ = 120 K convoluted with the energy resolution. The definition of the CDW-gap size $\Delta$ is shown in **e**. **k**, A set of EDCs at $T$ = 20 K extracted at the **k** points where the leading edge shows a smallest shift relative to $E_F$ along each cut (minimum gap locus). **l**, EDCs near the Γ and M points (green and red circles, respectively) together with the result of numerical fitting (light blue curve) assuming multiple Lorentzian peaks (black, green, and purple curves) multiplied by Fermi-Dirac distribution function (see Supplementary note 4 for details). Black dots show the peak position of the Sb-$p_z$ band near the Γ point and the SP1 band near the M point. **m**, Plot of estimated energy gap against $|\Delta k_x|$. **n** $h\nu$ dependence of the EDCs at the minimum gap locus of the M K cut in the surface Brillouin zone. **o**, Gap size plotted against $k_z$ estimated from the peak position of EDCs in **n**. Error bars in **m** and **o** reflect standard deviations in the numerically simulated $E_F$ position and peak position of EDCs.

To investigate how the 3D nature of CDW manifests itself in the CDW-gap properties, we have selected the $\overline{MK}$ cut at which the large CDW gap opens on the SP1 band, and carried out $h\nu$-dependent ARPES measurements. As shown in Fig. 2n, the EDC at $h\nu = 120$ eV corresponding to the M point ($k_z = 0$) displays a hump at $E_B \sim 80$ meV due to the CDW-gap opening. With decreasing $h\nu$, the hump systematically moves toward $E_F$, approaches closest to $E_F$ at $h\nu = 105$ eV (corresponding to the L point; $k_z = \pi$), and then disperses back toward the higher $E_B$ on further lowering $h\nu$. This indicates the $k_z$-dependent CDW gap, as summarized in Fig. 2o (also see Supplementary Figs. 5 and 6 for details in the estimation of the CDW-gap size). These results indicate that the 3D nature of CDW in $KV_3Sb_5$ shows up as a strong $k_z$-dependent CDW gap of the SP1 band. This is reasonable because the saddle point in the normal state gradually departs from $E_F$ on approaching the L point from the M point [29, 43, 44, 54], and consequently the CDW gap in the below-$E_F$ side becomes smaller on approaching the L point.

It is noted that, although the $k_z$ resolution in the present study ($\sim 0.18$ Å$^{-1}$; see Method) is about twice larger than that in soft x-ray ARPES [63], a complete loss of the resolution in $k_z$ is avoided and it is possible to perform ARPES measurements by selecting $k_z$ to some extent. When $h\nu$ is set to probe $k_z = 0$ or $\pi$, the electronic states at around $k_z = 0$ or $\pi$ are dominantly probed although the electronic states within $\pm 0.25$ $\pi/c$ centered at $k_z = 0$ or $\pi$ are partially involved. This condition, which is similar to the previous vacuum-ultraviolet ARPES study on the $k_z$-dependent gap [64], should ensure the validity of our qualitative argument that the gap at around $k_z = 0$ is larger than that at around $k_z = \pi$.

**Comparison between ARPES results and first-principles calculations**

A next important step is to pin down the structural distortion responsible for the FS reconstruction and CDW gap. For this sake, we have carried out first-principles band-structure calculations for the CDW phase of $KV_3Sb_5$ by assuming the SoD or TrH distortion. Figures 3a and 3b compare the calculated band dispersions along the ΓMKΓ line unfolded with respect to the original Brillouin zone for the SoD and TrH models, respectively. Structural parameters for each model were determined so that the downward shift of the SP1 band at the M point with respect to the undistorted one becomes similar to the experimental value of ~80 meV (actual structural parameters are listed in Supplementary Table 1); this comparison that properly stands on the experimental data would be meaningful because the calculations with lattice relaxation do not always reflect correct band parameters (note that the calculations with fully relaxed lattice parameters underestimate the CDW-gap size in the present case, as shown in Supplementary Fig. 7). One can see from Figs. 3a and 3b several similarities between the two calculations. For example, both calculations predict main bands with a strong intensity showing a broad correspondence with the original bands in the normal state (dashed green curves) as well as several weak sub-bands associated with the CDW-induced band folding. Thus, in order to distinguish an appropriate model that reproduces the experimental data, it may be useful to look at the ($E$, **k**) region where two models show a critically different spectral behavior, rather than to examine the overall agreement/disagreement of the valence-band dispersion. We found that the prominent difference in the calculated band structure is seen around the M point near $E_F$. This is reasonable because the proximity of the saddle point to $E_F$ plays a crucial role to reduce the total energy of system and thereby the band

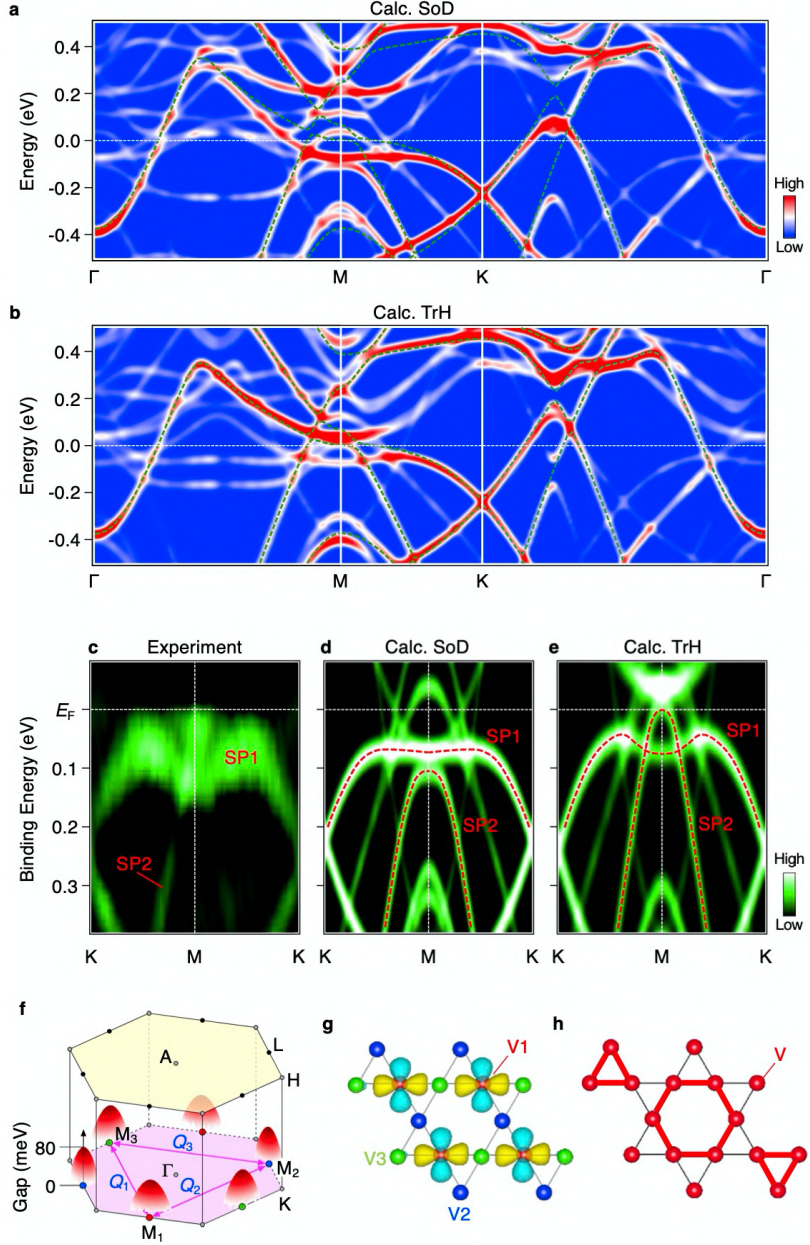

**FIG. 3. Calculated band structure and distinction of structural distortions. a, b** Calculated band dispersions along the ΓMKΓ cut for the SoD and TrH distortions, respectively. Calculated band dispersions for the undistorted 1×1 phase is shown by dashed green curves. Calculations have been performed with the 2×2×1 superlattice because the influence from the change of unit-cell (doubling or quadrupling) along the $c$ axis was not observed in the ARPES data, as seen from the $2\pi c^{-1}$ periodicity of the CDW-gap anisotropy along $k_z$ in Fig. 2**o**. **c**, ARPES intensity at $T$ = 20 K measured along the MK cut at $h\nu$ = 114 eV. **d, e** Calculated band dispersions along the MK cut for the SoD and TrH models, respectively. Red dashed curves are a guide for the eyes to trace the SP1 and SP2 bands. **f**, Schematics of **k**-dependent CDW gap on the SP1 band, together with the inter-saddle-point scattering vectors $Q_1$-$Q_3$ and three M points, $M_1$-$M_3$. **g**, Kagome lattice of V atoms with three different sublattices ($V_1$-$V_3$) shown with different coloring (red, blue, and green spheres, respectively). Calculated Wannier orbital for the SP1 band at one of the M points, which selectively places the V1 sublattice, is also indicated. **h**, Kagome lattice under the TrH distortion in which longer and shorter V-V bonds are highlighted by thick red and thin black lines, respectively.

structure associated with the saddle point is expected to be sensitive to the type of distortions.

To specify the appropriate model, we have chosen the MK cut in which the energy dispersion of SP1 and SP2 bands is well visible in the experiment. The ARPES-derived band dispersion along the MK cut at $T = 20$ K in Fig. 3c signifies an M-shaped structure below $E_F$ due to the SP1 band. The M-shaped feature does not exist in the calculated band dispersion in the normal state (green curves in Fig. 3a and 3b) because the SP1 band is associated with the V-$3d_{x^2-y^2}$-derived saddle point which is strongly modified by the **k**-dependent CDW gap as discussed in Figs. 2c-2f. One can recognize in Fig. 3c another rapidly dispersive Λ-shaped saddle-point band with the V-$3d_{xz/yz}$ character (SP2 band). As mentioned above, the top of this band stays near $E_F$ (within 20 meV of $E_F$; see Supplementary Fig. 8) and hence crosses the SP1 band slightly away from the M point. As shown in Figs. 3d and 3e, while the calculated band structures for both the SoD and TrH models qualitatively reproduce a gap opening of the SP1 band, the TrH model shows a better agreement on the M-shaped dispersion observed in the experiment. Further, a difference between the two models is observed in the behavior of the SP2 band; it sinks below the SP1 band at the M point in the SoD model whereas it penetrates the SP1 band in the TrH model. Such a difference is also seen in the DFT calculations performed with a different code [36], suggesting a tendency that the experimental data in Fig. 3c is more likely reproduced by the TrH distortion. Therefore, the TrH distortion may actually take place in the CDW phase of $KV_3Sb_5$. To obtain a decisive conclusion on this point, additional effects which are not taken into account in the present study, e.g., band renormalization and resultant change in the gap value as well as more complex structural distortions, should be investigated.

The 3D CDW gap and the possible TrH structural distortion discussed above put a strong constraint on the mechanism of CDW and superconductivity, as well as on the possible topological nature of superconducting state. In general, the electronic energy gain associated with the CDW is governed by the **k** region where a large CDW gap opens at $E_F$. The observed strong **k**-dependent CDW gap, which takes the maximum at the M point as a function of both $k_z$ and $k_{//}$, strongly suggests that the proximity to $E_F$ of the saddle-point band (SP1 band) with the V-$3d_{x^2-y^2}$ character plays a major role in lowering the electron kinetic energy. In addition, the absence of a clear folding of the $k_z$-dependent CDW gap suggests that, despite the existence of unit-cell doubling or quadrupling along the *c* axis [33, 35, 36], the V electrons in the kagome plane do not feel so strongly the superlattice potential. Therefore, the electron scattering by the in-plane *Q* vector connecting two different M points is a leading factor to stabilize the CDW in $KV_3Sb_5$ (Fig. 3f). While such *Q* vector is unidirectional, the existence of three equivalent *Q* vectors ($Q_1$-$Q_3$ in Fig. 3f) connecting three different M points ($M_1$-$M_3$) in the Brillouin zone leads to the 3*Q*-CDW state that satisfies the 2×2 periodicity. Besides the 3*Q*-CDW, the possible TrH distortion can be also explained by the bond-order state. Namely, the SP1 band at the $M_1$-$M_3$ points in the normal state is predominantly occupied by electrons in three different sublattices composed of V1-V3 atoms (see sublattice-selective distribution of the Wannier orbital for the SP1 band in Fig. 3g), and the inter-saddle-point electron scattering via the $Q_1$-$Q_3$ vectors enhances the sublattice interference effect [26]. This results in shortening of neighboring V-V bonds indicated by thick red lines in Fig. 3h (note that other V-V bonds indicated by thin black lines become longer), leading to tiling of hexagonal and triangle bond patterns consistent with the TrH distortion [55, 65]. Also, the lattice distortion responsible for the 2×2 CDW is critically important to pin down the superconducting pairing symmetry because the location of gap nodes and the topological nature of superconductivity are directly linked to the type of structural distortion in the ground state [66]. The present study further suggests that the CDW-induced reconstructed FS, but not the normal-state FS, must be considered to construct a microscopic theory of superconductivity in AV$_3$Sb$_5$.

**METHOD**

**ARPES measurements**

High-quality single crystals of KV$_3$Sb$_5$ were synthesized by the self-flux method [40]. Photon-energy-tunable vacuum ultraviolet ARPES measurements were performed with the MBS-A1 analyzer at BL5U in UVSOR. We used linearly polarized light of 80–140 eV. The energy resolution was set to be 10-30 meV. The angular resolution was set to be 0.3º, which corresponds to the resolution in the momentum parallel to the kagome plane ($k_{//}$) of 0.03-0.04 Å$^{-1}$. The resolution in the momentum perpendicular to the kagome plane ($k_z$) is estimated to be ~0.18 Å$^{-1}$, which corresponds to a half of the Γ-A distance in the bulk Brillouin zone (~0.5 π $c^{-1}$) [note that, according to the universal curve [67], the electron mean-free path λ at the present $h\nu$ range (~100 eV) is ~5.5 Å, leading to the $k_z$ broadening δ$k_z$ expressed as λ$^{-1}$ of ~0.18 Å$^{-1}$]. Samples were cleaved *in situ* along the (0001) plane of the hexagonal crystal in an ultrahigh vacuum of 1×10$^{-10}$ Torr, and kept at $T$ = 20 or 120 K during the measurements.

**Band calculations**

First-principles band-structure calculations were carried out by using the Quantum Espresso code package [68] with generalized gradient approximation [69]. Spin-orbit coupling and D3 corrections were included in the calculations unless otherwise stated.

The plane-wave cutoff energy and the *k*-point mesh were set to be 50 Ry and 11×11×5, respectively. Supercell calculations were carried out on a 2×2×1 supercell with a 5×5×5 *k* mesh. The unfolding of calculated bands was performed using the BandUP code [70]. Wannier orbital was calculated by using the Wannier90 code [71].

## DATA AVAILABILITY

The data sets generated/analyzed during the current study are included in the published article and its Supplementary Information file. The numerical data sets are available from the corresponding author on reasonable request.

## CODE AVAILABILITY

Details on the numerical fittings and band-structure calculations are available from the corresponding author on reasonable request.

**Acknowledgments**

This work was supported by JST-CREST (No. JPMJCR18T1), JST-PRESTO (No. JPMJPR18L7), Grant-in-Aid for Scientific Research (JSPS KAKENHI Grant Numbers JP21H04435 and JP20H01847), and UVSOR (Proposal number: 21-658 and 21-679). The work at Beijing was supported by the National Key R&D Program of China Grant No. 2020YFA0308800), the Natural Science Foundation of China (Grants No. 92065109, No. 11734003, and No. 12061131002), the Beijing Natural Science Foundation (Grant No. Z190006), and the Beijing Institute of Technology (BIT) Research Fund Program for Young Scholars (Grant No. 3180012222011). T. Kato and T. Kawakami acknowledge support from GP-Spin. Z.W. thanks the Analysis & Testing Center at BIT for assistance in facility support.

**AUTHOR CONTRIBUTIONS**

The work was planned and proceeded by discussion among T. Kato, K.N. and T.S. T. Kato, T. Kawakami, K.N., A.M., K.T., T.T., and T.S. performed the ARPES measurements. Y.L. and Z.W. carried out the growth and characterization of crystals. T.

Kawakami, M.L., and Y.Y. carried out the band-structure calculations. T. Kato, K.N., and T.S. finalized the manuscript with inputs from all the authors.

## COMPETING INTERESTS

The authors declare no competing interests

## ADDITIONAL INFORMATION

**Supplementary information** is available for this paper at http://...

**Correspondence** and requests for materials should be addressed to K. N. (k.nakayama@arpes.phys.tohoku.ac.jp), Z. W. (zhiweiwang@bit.edu.cn), and T. S. (e-mail: t-sato@arpes.phys.tohoku.ac.jp).